\PassOptionsToPackage{unicode}{hyperref}
\PassOptionsToPackage{hyphens}{url}
\PassOptionsToPackage{dvipsnames,svgnames,x11names}{xcolor}
\documentclass[
]{article}
\usepackage{xcolor}
\usepackage[margin=1in]{geometry}
\usepackage{amsmath,amssymb}
\usepackage{iftex}
\ifPDFTeX
  \usepackage[T1]{fontenc}
  \usepackage[utf8]{inputenc}
  \usepackage{textcomp} 
\else 
  \usepackage{unicode-math} 
  \defaultfontfeatures{Scale=MatchLowercase}
  \defaultfontfeatures[\rmfamily]{Ligatures=TeX,Scale=1}
\fi
\usepackage{lmodern}
\ifPDFTeX\else
\fi
\IfFileExists{upquote.sty}{\usepackage{upquote}}{}
\IfFileExists{microtype.sty}{
  \usepackage[]{microtype}
  \UseMicrotypeSet[protrusion]{basicmath} 
}{}
\makeatletter
\@ifundefined{KOMAClassName}{
  \IfFileExists{parskip.sty}{%
    \usepackage{parskip}
  }{
    \setlength{\parindent}{0pt}
    \setlength{\parskip}{6pt plus 2pt minus 1pt}}
}{
  \KOMAoptions{parskip=half}}
\makeatother
\makeatletter
\ifx\paragraph\undefined\else
  \let\oldparagraph\paragraph
  \renewcommand{\paragraph}{
    \@ifstar
      \xxxParagraphStar
      \xxxParagraphNoStar
  }
  \newcommand{\xxxParagraphStar}[1]{\oldparagraph*{#1}\mbox{}}
  \newcommand{\xxxParagraphNoStar}[1]{\oldparagraph{#1}\mbox{}}
\fi
\ifx\subparagraph\undefined\else
  \let\oldsubparagraph\subparagraph
  \renewcommand{\subparagraph}{
    \@ifstar
      \xxxSubParagraphStar
      \xxxSubParagraphNoStar
  }
  \newcommand{\xxxSubParagraphStar}[1]{\oldsubparagraph*{#1}\mbox{}}
  \newcommand{\xxxSubParagraphNoStar}[1]{\oldsubparagraph{#1}\mbox{}}
\fi
\makeatother

\usepackage{longtable,booktabs,array}
\usepackage{calc} 
\usepackage{etoolbox}
\makeatletter
\patchcmd\longtable{\par}{\if@noskipsec\mbox{}\fi\par}{}{}
\makeatother
\IfFileExists{footnotehyper.sty}{\usepackage{footnotehyper}}{\usepackage{footnote}}
\makesavenoteenv{longtable}
\usepackage{graphicx}
\makeatletter
\newsavebox\pandoc@box
\newcommand*\pandocbounded[1]{
  \sbox\pandoc@box{#1}%
  \Gscale@div\@tempa{\textheight}{\dimexpr\ht\pandoc@box+\dp\pandoc@box\relax}%
  \Gscale@div\@tempb{\linewidth}{\wd\pandoc@box}%
  \ifdim\@tempb\p@<\@tempa\p@\let\@tempa\@tempb\fi
  \ifdim\@tempa\p@<\p@\scalebox{\@tempa}{\usebox\pandoc@box}%
  \else\usebox{\pandoc@box}%
  \fi%
}
\def\fps@figure{htbp}
\makeatother

\usepackage[]{natbib}
\usepackage{graphicx} 
\usepackage{fullpage}
\usepackage{authblk}
\newcommand\loglik{\lambda}
\newcommand\fproc{f_{\mathrm{proc}}}
\newcommand\fmeas{f_{\mathrm{meas}}}
\newcommand\sproc{s_{\mathrm{proc}}}
\newcommand\smeas{s_{\mathrm{meas}}}

\newcommand\fprocDiscrete{\mathrm{discrete}}
\newcommand\fprocEuler{\mathrm{Euler}}
\newcommand\fmeasBinomial{\mathrm{binomial}}
\newcommand\fmeasGaussian{\mathrm{Gaussian}}

\newcommand\sprocDiscrete{\mathrm{discrete}}
\newcommand\sprocEuler{\mathrm{Euler}}
\newcommand\smeasBinomial{\mathrm{binomial}}
\newcommand\smeasGaussian{\mathrm{Gaussian}}

\newcommand{\PALL}{\ensuremath{\mathrm{PAL}_L}}
\newcommand{\PALV}{\ensuremath{\mathrm{PAL}_V}}
\usepackage{multirow}
\usepackage[dvipsnames]{xcolor}

\makeatletter
\@ifpackageloaded{caption}{}{\usepackage{caption}}
\AtBeginDocument{%
\ifdefined\contentsname
  \renewcommand*\contentsname{Table of contents}
\else
  \newcommand\contentsname{Table of contents}
\fi
\ifdefined\listfigurename
  \renewcommand*\listfigurename{List of Figures}
\else
  \newcommand\listfigurename{List of Figures}
\fi
\ifdefined\listtablename
  \renewcommand*\listtablename{List of Tables}
\else
  \newcommand\listtablename{List of Tables}
\fi
\ifdefined\figurename
  \renewcommand*\figurename{Figure}
\else
  \newcommand\figurename{Figure}
\fi
\ifdefined\tablename
  \renewcommand*\tablename{Table}
\else
  \newcommand\tablename{Table}
\fi
}
\@ifpackageloaded{float}{}{\usepackage{float}}
\floatstyle{ruled}
\@ifundefined{c@chapter}{\newfloat{codelisting}{h}{lop}}{\newfloat{codelisting}{h}{lop}[chapter]}
\floatname{codelisting}{Listing}

\makeatother
\makeatletter
\@ifpackageloaded{caption}{}{\usepackage{caption}}
\@ifpackageloaded{subcaption}{}{\usepackage{subcaption}}
\makeatother
\usepackage{bookmark}
\IfFileExists{xurl.sty}{\usepackage{xurl}}{} 
\hypersetup{
  pdfauthor={Kunyang He, Yize Hao and Edward L. Ionides; University of Michigan, Ann Arbor},
  colorlinks=true,
  linkcolor={blue},
  filecolor={Maroon},
  citecolor={Blue},
  urlcolor={Blue},
  pdfcreator={LaTeX via pandoc}}

\title{Poisson Approximate Likelihood versus the block particle filter
for a spatiotemporal measles model}
\author{Kunyang He, Yize Hao and Edward L. Ionides \and University of
Michigan, Ann Arbor}
\date{}
\begin{document}
\maketitle
\begin{abstract}
Filtering algorithms for high-dimensional nonlinear non-Gaussian
partially observed stochastic processes provide access to the likelihood
function and hence enable likelihood-based or Bayesian inference for
this methodologically challenging class of models. A novel Poisson
approximate likelihood (PAL) filter was introduced by Whitehouse et
al.~(2023). PAL employs a Poisson approximation to conditional
densities, offering a fast approximation to the likelihood function for
a certain subset of partially observed Markov process models. PAL was
demonstrated on an epidemiological metapopulation model for measles,
specifically, a spatiotemporal model for disease transmission within and
between cities. At face value, Table~3 of Whitehouse et al.~(2023)
suggests that PAL considerably out-performs previous analysis as well as
an ARMA benchmark model. We show that PAL does not outperform a block
particle filter and that the lookahead component of PAL was implemented
in a way that introduces substantial positive bias in the log-likelihood
estimates. Therefore, the results of Table~3 of Whitehouse et al.~(2023)
do not accurately represent the true capabilities of PAL.
\end{abstract}

\section{Introduction}
\label{sec:intro}

Investigations of the metapopulation dynamics of measles (i.e., studying
how measles infection moves within and between collections of spatially
distinct populations) have motivated various methodological innovations
for inference on high-dimensional partially observed stochastic
processes \citep{xia04,park20,ionides23-jasa}. The analysis by
\citet{whitehouse23} (henceforth, WWR) provides a new approach to
model-based inference on population dynamics via the Poisson approximate
likelihood (PAL) filtering algorithm. WWR claimed impressive results on
both a low-dimensional rotavirus transmission model and a
high-dimensional measles model. On close inspection, the rotavirus
results turned out to be overstated \citep{hao24-arxiv} leading to a
published correction \citep{whitehouse25-correction}. However, the
spatiotemporal measles results were unaffected by that correction, and
our present purpose is to revisit this example.

In Section~\ref{sec:numerics} we show by direct numerical
experimentation that the WWR implementation of PAL has substantial
positive Monte Carlo bias in the log-likelihood estimate, and so the
comparison of log-likelihood values to support the use of the method is
flawed. We establish and interpret two main lines of evidence: (i) on
simulated data, WWR's methodology can lead to log-likelihood estimates
considerably higher than the known value of the log-likelihood; (ii), on
simulated and real data, applying WWR's algorithm with increasing Monte
Carlo effort leads to decreasing log-likelihood estimates. While
investigating these phenomena, we show that a widely applicable block
particle filter (BPF) is adequate on this problem. In
Section~\ref{sec:theory} we explain theoretically how the positive Monte
Carlo bias for PAL in the measles example arises as a result of the
lookahead mechanism included in the implementation of PAL that WWR used
for this model. The lookahead mechanism was not used by WWR for the
rotavirus analysis since a basic version of PAL was sufficient for that
lower-dimensional example. Section~\ref{sec:conclusion} is a concluding
discussion.

For our current purposes, we do not have to delve into the details of
the measles data and model, so we provide only a brief overview. The
data are measles case counts aggregated over 2-week intervals for forty
of the largest towns in England and Wales, from 1949 to 1964. The data
and the model are derived from \citet{park20}, who built on a long
tradition of models described therein. Recently, weekly data for more
towns have become publicly available \citep{korevaar20}, but we limit
ourselves to the data used by WWR. The latent process model describes an
integer count of infected, susceptible and recovered individuals in each
town. The rate of disease transmission within cities follows widely used
epidemiological equations. Transmission between pairs of cities follows
a power law, diminishing with distance between the cities. This is known
as a gravity model. Overdispersion for the latent dynamics is achieved
by placing multiplicative gamma white noise on the transmission rate.
The measurement model is a discretized Gaussian approximation to an
overdispersed binomial \citep{park20} or Gaussian noise on a binomial
rate (WWR). \citet{park20} used a particle filter known as a guided
intermediate resampling filter (GIRF). Recently, BPF has been shown to
have good performance on this class of models
\citep{ionides23-jasa,ionides24-sinica,ning23}. Therefore, we compare
PAL with BPF.

\section{Numerical experiments for PAL and BPF}
\label{sec:numerics}

There are many possible numerical experiments that could be conducted to
compare filters on spatiotemporal measles models. Here, we choose
experiments to investigate two specific hypotheses:

\begin{enumerate}
\item[H1] The lookahead version of the Poisson approximate likelihood estimator of WWR, which we call $\PALL$, can have substantial positive Monte Carlo bias on its Monte Carlo log-likelihood estimate. This occurs in the spatiotemporal example of WWR.

\item[H2] The Monte Carlo bias in the $\PALL$ log-likelihood estimate scales approximately linearly with the number of spatial units.
\end{enumerate}

We consider probabilistic filtering algorithms that are defined in the
context of a model, its model parameters, and additional algorithmic
parameters. Additionally, we require data, and this can either be the
real historical measles data or can be simulated from another model that
may or may not be the same model with the same parameters as used for
the filter. We also have a choice of how many spatial units to include,
with each unit being one UK town in the measles example. The
experimental variables, and the list of values we consider for them, are
summarized in Table~\ref{tbl-description} and further described below.

A central part of our reasoning is that a probabilistic forecasting
filter (i.e., one that solves the one-step prediction problem without
looking ahead to future data) cannot, on average, obtain higher
log-likelihood than the exact prediction distribution, when the average
is taken across many datasets generated by the exact model. This is a
restatement of the well-known fact that log-likelihood is a proper
scoring rule \citep{gneiting07}. To apply this property, we must work
with simulated data so that the true generating model is known. In high
dimensions, it is generally not possible to calculate the exact
prediction distribution to within arbitrarily small error. Monte Carlo
methods offering provably consistent estimation of this distribution, in
a limit as the Monte Carlo effort increases, have intolerable
Monte~Carlo error. That is the reason why approximation algorithms such
as \(\PALL\) are being invented. There are two special situations where
we can establish accurately the true log-likelihood for the
spatiotemporal measles models of interest: (i) when the number of
spatial units is very small; (ii) when there is no spatial coupling, so
the filtering problem can be solved independently for each single unit.
In both these cases, a basic particle filter provides the desired,
essentially exact, log-likelihood estimate. The basic particle filter is
consistent and unbiased for the likelihood \citep{delmoral04} and so,
when its estimates have low empirical variance, it provides the required
ground truth. In practice, order \(10^5\) particles give a highly
accurate log-likelihood for one unit, but the strong sensitivity of the
particle filter to the curse of dimensionality \citep{bengtsson08} means
that quantifiably exact estimates rapidly become unfeasible. Therefore,
we consider two choices of size for the system, \(U=1\) and \(U=40\),
with the latter being the size of the system tested by WWR. To allow for
the study of systems without coupling (i.e., measles transmission
between cities), we consider a model variation, \(C_2\), for which there
is no direct movement of infection between cities and instead there is a
constant background rate of importation of infection. By contrast,
\(C_1\) is a specification with coupling following a gravity model, as
used by WWR.

One approach to address the limitations of particle filters for
high-dimensional systems is to take advantage of the possibility to
design the filter's prediction distribution for the \(n\)th observation,
at time \(t_n\), to build a prediction distribution that incorporates
data occurring at, or subsequent to, time \(t_n\). Such lookahead
filters cannot be implemented for forecasting, but can be used for
likelihood evaluation. This is the approach adopted by WWR's lookahead
filter, \(\PALL\), which incorporates a particle filter to address
overdispersion in the data.

There is no mathematical theorem prohibiting a lookahead filter
estimating a higher likelihood than the truth, and in an extreme case
the lookahead filter could just assert a one-step prediction
distribution with all its mass on the actual data. Lookahead filters
therefore need careful theoretical guarantees if we want to use a high
likelihood estimate as evidence for both the success of the filter and
(when doing data analysis) evidence supporting the model used to
construct the filter. We will investigate the theory behind \(\PALL\)
later, in Section~\ref{sec:theory}, but for now we just identify the
potential hazard.

\begin{table}

\caption{\label{tbl-description}Variables for the numerical experiments and their set of values.}

\centering{

\begin{tabular}{lllll}
Variable & Description & Value 1 & Value 2 & Value 3
\\
\hline
$F$ & filter algorithm & \PALL & \PALV & BPF
\\
$J$ & number of particles & $J_1=5\times 10^3$ &  $J_2=10^5$ &
\\
$U$ & number of spatial units & $U_1=1$ &  $U_2=40$ &
\\
$f_C$ & spatiotemporal coupling for filter & $C_1 =(g\neq 0, \iota = 0)$ & $C_2=(g=0, \iota\neq 0)$ &
\\
$\fproc$ & process model for filter & $\fprocEuler$ & $\fprocDiscrete$ &
\\
$\fmeas$ & measurement model for filter & $\fmeasBinomial$ & $\fmeasGaussian$ &
\\
$f_{\theta}$ & parameter for filter & $\hat\theta^*_{BPF}$ & $\hat\theta_{PAL}$ &
\\
$s_C$ & spatiotemporal coupling for simulation & $C_1 =(g\neq 0, \iota = 0)$ & $C_2=(g=0, \iota\neq 0)$ &
\\
$\sproc$ & process model for simulation & $\sprocEuler$ & $\sprocDiscrete$ &
\\
$\smeas$ & measurement model for simulation & $\smeasBinomial$ & $\smeasGaussian$ &
\\
$s_{\theta}$ & parameter for simulation & $\hat\theta^*_{BPF}$ & $\hat\theta_{PAL}$ &
\\
\hline
\end{tabular}

}

\end{table}%

The experimental variables listed in Table~\ref{tbl-description} are now
described in more detail:

\begin{itemize}
\item[F] {\bf The filtering algorithm}. 
{\PALL} is the lookahead PAL sequential Monte Carlo filter of WWR, and {\PALV} is the plain, so-called vanilla, implementation. 
BPF is the block particle filter of \citet{rebeschini15} implemented as \texttt{bpfilter} in spatPomp \citep{asfaw24}.
Likelihood optimization for {\PALL} and {\PALV} is conducted using stochastic gradient descent and automatic differentiation, using the implementation by WWR. 
Likelihood optimization for BPF is conducted using the iterated BPF algorithm \citep{ning23,ionides24-sinica} implemented as \texttt{ibpf} in spatPomp.
For $U=1$ (and for $U=40$ with $g=0$) BPF is identical to a basic particle filter (or collection of independent particle filters). 
For simplicity, we use \texttt{bpfilter} from the spatPomp package even when \texttt{pfilter} from the pomp package is equivalent.

\item[$f_C$]  {\bf Spatiotemporal coupling for the filter model}.
The choice $C=C_1$ corresponds to the coupling used by WWR, with spatial movement of infection ($g\neq 0$) and no background immigration of infection from outside the study system ($\iota=0$). 
In order to test the methods on a high-dimensional system for which the true likelihood is known to a good degree of accuracy, we also consider setting $C_2$, without coupling ($g=0$) and with compensating immigration to prevent permanent extinction of measles in small towns ($\iota\neq 0$).
When $U=U_1=1$, we use the largest city, London, for which stochastic extinctions are very unlikely. 
Note that, when $U=1$, the value of $g$ becomes irrelevant.

\item[$\fproc$]  {\bf Latent process transition model for the filter}.
For $\fproc=\fprocDiscrete$,  a single gamma-distributed dynamic noise variable is chosen for each observation interval.
This is the choice made by WWR.
For $\fproc=\fprocEuler$, independent gamma noise variables are included in each Euler time step, so that the limit of the process model (as the Euler time step decreases) corresponds to a continuous-time over-dispersed Markov chain \citep{breto11}. 
Both $\fproc=\fprocDiscrete$ and $\fproc=\fprocEuler$ are implemented with a step of $1/2$ week for the multinomial transitions conditional on the gamma noise. 

\item[$\fmeas$]  {\bf Measurement model for the filter}. The choice of WWR is $\fmeas=\fmeasBinomial$, corresponding to binomial measurements with truncated multiplicative Gaussian noise on the reporting rate, i.e., the expected fraction of infections that are reported. 
The basic PAL algorithm requires a binomial measurement model, but the SMC-PAL extension permits noise on the measurement probability.
We also consider $\fmeas=\fmeasGaussian$, corresponding to a discretized Gaussian measurement model. 
This choice leads to a measurement model that can directly be evaluated, without costly Monte Carlo calculation, assisting with efficient Monte Carlo inference.

\item[$f_{\theta}$]  {\bf Model parameter vector for the filter}. 
{\PALL} and {\PALV} are evaluated at optimized parameter vectors for each data set. 
For PAL, there is generally no true POMP model for which PAL is an exact filter.
However, we give PAL a reasonable chance to show its capabilities by optimizing it using the code provided by WWR.
When $U>1$, we must decide which parameter are shared between units and which are unit-specific.
For $\PALL$ and $\PALV$, we used case A from WWR, for which parameters are shared between units.
This ran much more quickly than their case C, and is sufficient to make our point.
$E_{12}$-$E_{15}$ use the parameter values published by WWR for their case A.
For $E_{11}$, we used case C.

\item[$\sproc$]  {\bf Process model for the simulation}. 
Always set to $\sproc=\sprocEuler$ since simulations were carried out using an implementation of the model in spatPomp. 

\item[$\smeas$]  {\bf Measurement model for the simulation}. 
Always set to $\smeas=\smeasGaussian$ since simulations were carried out using an implementation of the model in spatPomp.

\item[$s_{\theta}$]  {\bf Model parameter vector for the simulated data}.
To have an essentially exact likelihood evaluation using BPF, we have $s_\theta=f_\theta$ for all BPF situations. 
The parameters were fixed at a value near the maximum likelihood estimate for the data.

\end{itemize}

We consider two software platforms for the experiments. All \(\PALL\)
and \(\PALV\) calculations were carried out using the Python code
provided by WWR, and all BPF calculations were carried out using the
spatPomp R package. Simulations were carried out using spatPomp, since
part of our reasoning (the use of the proper scoring rule property of
log-likelihood) depends critically on the particle filter calculations
being carried out with the data drawn from the model assumed by the
filter. We present results for a single simulation for the experimental
treatments with simulated data. That decision simplifies the
experimental design and permits the computational effort to focus on a
few direct comparisons between the methods on a small number of
simulated datasets. The experiments carried our are described in
Table~\ref{tbl-treatments}.

\begin{table}

\caption{\label{tbl-treatments}Combinations of variable values used for each experiment, $E_k$, $k=1\dots 15$. The simulation settings, $s_C$, $\sproc$, $\smeas$ and $s_\theta$, are applicable only when we filter simulated data rather than real data.}

\centering{

\centering
\begin{tabular}{llllllllllll}
E & F & J & U & 
  $f_C$ & $\fproc$ &  $\fmeas$ & $f_{\theta}$ &
  $s_C$ & $\sproc$ &  $\smeas$ & $s_{\theta}$ 
\\
\hline
$E_1$ &  BPF & $J_2$ & $U_1$ & 
  $C_1$ & $\fprocEuler$ & $\fmeasGaussian$ & $\hat\theta^*_{BPF}$ &
  $C_1$ & $\fprocEuler$ & $\fmeasGaussian$ & $\hat\theta^*_{BPF}$
\\
$E_2$ &  \PALV & $J_1$ & $U_1$ & 
  $C_1$ & $\fprocDiscrete$ & $\fmeasBinomial$ & $\hat\theta_{PAL}$ &
  $C_1$ & $\fprocEuler$ & $\fmeasGaussian$ & $\hat\theta^*_{BPF}$
\\
$E_3$ &  \PALV & $J_2$ & $U_1$ & 
  $C_1$ & $\fprocDiscrete$ & $\fmeasBinomial$ & $\hat\theta_{PAL}$ &
  $C_1$ & $\fprocEuler$ & $\fmeasGaussian$ & $\hat\theta^*_{BPF}$
\\
$E_4$ &  \PALL & $J_1$ & $U_1$ & 
  $C_1$ & $\fprocDiscrete$ & $\fmeasBinomial$ & $\hat\theta_{PAL}$ &
  $C_1$ & $\fprocEuler$ & $\fmeasGaussian$ & $\hat\theta^*_{BPF}$
\\
$E_5$ &  \PALL & $J_2$ & $U_1$ & 
  $C_1$ & $\fprocDiscrete$ & $\fmeasBinomial$ & $\hat\theta_{PAL}$ &
  $C_1$ & $\fprocEuler$ & $\fmeasGaussian$ & $\hat\theta^*_{BPF}$
\\
$E_6$ &  \PALL & $J_1$ & $U_1$ & 
  $C_1$ & $\fprocDiscrete$ & $\fmeasBinomial$ & $\hat\theta_{PAL}$ &
  \multicolumn{4}{c}{\hfill --------------- \hfill data \hfill --------------- \hfill}
\\
$E_7$ &  BPF & $J_1$ & $U_1$ & 
  $C_1$ & $\fprocEuler$ & $\fmeasGaussian$ & $\hat\theta^*_{BPF}$ &
  \multicolumn{4}{c}{\hfill --------------- \hfill data \hfill --------------- \hfill}
\\
$E_8$ &  BPF & $J_2$ & $U_2$ & 
  $C_2$ & $\fprocEuler$ & $\fmeasGaussian$ & $\hat\theta^*_{BPF}$ &
  $C_2$ & $\fprocEuler$ & $\fmeasGaussian$ & $\hat\theta^*_{BPF}$
\\
$E_9$ &  \PALV & $J_1$ & $U_2$ & 
  $C_1$ & $\fprocDiscrete$ & $\fmeasBinomial$ & $\hat\theta_{PAL}$ &
  $C_2$ & $\fprocEuler$ & $\fmeasGaussian$ & $\hat\theta^*_{BPF}$
\\
$E_{10}$ &  \PALL & $J_1$ & $U_2$ & 
  $C_1$ & $\fprocDiscrete$ & $\fmeasBinomial$ & $\hat\theta_{PAL}$ &
  $C_2$ & $\fprocEuler$ & $\fmeasGaussian$ & $\hat\theta^*_{BPF}$
\\
$E_{11}$ &  BPF & $J_2$ & $U_2$ & 
  $C_1$ & $\fprocEuler$ & $\fmeasGaussian$ & $\hat\theta^*_{BPF}$ &
  \multicolumn{4}{c}{\hfill --------------- \hfill data \hfill --------------- \hfill}
\\
$E_{12}$ &  \PALV & $J_1$ & $U_2$ & 
  $C_1$ & $\fprocDiscrete$ & $\fmeasBinomial$ & $\hat\theta_{PAL}$ &
  \multicolumn{4}{c}{\hfill --------------- \hfill data \hfill --------------- \hfill}
\\
$E_{13}$ &  \PALV & $J_2$ & $U_2$ & 
  $C_1$ & $\fprocDiscrete$ & $\fmeasBinomial$ & $\hat\theta_{PAL}$ &
  \multicolumn{4}{c}{\hfill --------------- \hfill data \hfill --------------- \hfill}
\\
$E_{14}$ &  \PALL & $J_1$ & $U_2$ & 
  $C_1$ & $\fprocDiscrete$ & $\fmeasBinomial$ & $\hat\theta_{PAL}$ &
  \multicolumn{4}{c}{\hfill --------------- \hfill data \hfill --------------- \hfill}
\\
$E_{15}$ &  \PALL & $J_2$ & $U_2$ & 
  $C_1$ & $\fprocDiscrete$ & $\fmeasBinomial$ & $\hat\theta_{PAL}$ &
  \multicolumn{4}{c}{\hfill --------------- \hfill data \hfill --------------- \hfill}
\\
\hline
 \end{tabular}

}

\end{table}%

Each experiment, \(E_k\), has two primary outcomes, a log-likelihood
estimate, \(\loglik_k\), and its standard error, \(\sigma_k\). These
results are tabulated in Table~\ref{tbl-method-comparison}, together
with benchmark log-likelihoods for a log-ARMA(2,1) model and a negative
binomial autoregressive model.

\begin{table}

\caption{\label{tbl-method-comparison}Log‐likelihood estimate, $\lambda$,  for each experiment described in Table\ 2.
Estimates derive from averaging 20 replicated Monte Carlo evaluations.
We averaged on a natural scale so that the basic particle filter estimate is unbiased.
The standard error, $\sigma$, is a jack-knife estimate implemented via the logmeanexp function in the pomp R package.
When $\sigma\gg 1$, this standard error is unreliable and we conclude only that the error is large.
ARMA gives the log-likelihood for an autoregressive moving average benchmark, and NegBinom is an autoregressive negative binomial benchmark.}

\centering{

\centering
\begin{tabular}{llcccrrrr}
\toprule
\textbf{$E$} & \textbf{$F$} & \textbf{$J$} & \textbf{$U$} & \textbf{$f_C$} &
$\lambda$ & $\sigma$ & ARMA & NegBinom \\
\midrule
$E_1$   & $\mathrm{BPF}$ & $J_2$ & $U_1$ & $C_1$ &
$-3133.06$  &
$0.03$  &
$-3190.08$   &
$-3248.91$   \\[2pt]

$E_2$   & $\PALV$        & $J_1$ & $U_1$ & $C_1$ &
$-3208.66$  &
$0.56$  &
$-3190.08$   &
$-3248.91$   \\[2pt]

$E_3$   & $\PALV$        & $J_2$ & $U_1$ & $C_1$ &
$-3206.26$  &
$0.22$  &
$-3190.08$   &
$-3248.91$   \\[2pt]

$E_4$   & $\PALL$        & $J_1$ & $U_1$ & $C_1$ &
$-2984.61$  &
$0.92$  &
$-3190.08$   &
$-3248.91$   \\[2pt]

$E_5$   & $\PALL$        & $J_2$ & $U_1$ & $C_1$ &
$-2993.40$  &
$0.45$  &
$-3190.08$   &
$-3248.91$   \\[2pt]

$E_6$   & $\PALL$        & $J_1$ & $U_1$ & $C_1$ &
$-2449.86$  &
$1.86$  &
$-2561.32$   &
$-2751.23$   \\[2pt]

$E_7$   & $\mathrm{BPF}$ & $J_1$ & $U_1$ & $C_1$ &
$-2501.15$  &
$0.32$  &
$-2561.32$   &
$-2751.23$   \\[2pt]

$E_8$   & $\mathrm{BPF}$ & $J_2$ & $U_2$ & $C_2$ &
$-79611.19$  &
$0.18$  &
$-82301.84$   &
$-84818.27$   \\[2pt]

$E_9$   & $\PALV$        & $J_1$ & $U_2$ & $C_1$ &
$-81445.26$  &
$7.78$  &
$-82301.84$   &
$-84818.27$   \\[2pt]

$E_{10}$ & $\PALL$       & $J_1$ & $U_2$ & $C_1$ &
$-78223.41$ &
$12.86$ &
$-82301.84$  &
$-84818.27$  \\[2pt]

$E_{11}$ & $\mathrm{BPF}$ & $J_2$ & $U_2$ & $C_1$ &
$-67761.82$ &
$0.78$ &
$-69168.10$  &
$-72401.20$  \\[2pt]

$E_{12}$ & $\PALV$       & $J_1$ & $U_2$ & $C_1$ &
$-74709.92$ &
$206.77$ &
$-69168.10$  &
$-72401.20$  \\[2pt]

$E_{13}$ & $\PALV$       & $J_2$ & $U_2$ & $C_1$ &
$-74137.62$ &
$4.23$ &
$-69168.10$  &
$-72401.20$  \\[2pt]

$E_{14}$ & $\PALL$       & $J_1$ & $U_2$ & $C_1$ &
$-64014.93$ &
$60.48$ &
$-69168.10$  &
$-72401.20$  \\[2pt]

$E_{15}$ & $\PALL$       & $J_2$ & $U_2$ & $C_1$ &
$-64848.76$ &
$2.25$ &
$-69168.10$  &
$-72401.20$  \\
\bottomrule
\end{tabular}

}

\end{table}%

Here, experiments \(E_1\)--\(E_5\) provide a computationally tractable
comparison using simulated data on a single unit. \(E_1\) provides a
ground truth for this particular model, a single-city SEIR model.
Comparing \(\lambda_2\) and \(\lambda_3\) with \(\lambda_1\), we see
that \(\PALV\), without the lookahead, performs as expected for an
approximate filter. On this relatively easy task, it produces stable
estimates, with log-likelihood values somewhat below the truth. However,
both \(E_2\) and \(E_3\) failed to outperform the ARMA benchmark in our
experiments, suggesting that PAL continues to face challenges when
dealing with complex SEIR models and overdispersion. Since \(\PALV\) is
filtering using a model that differs slightly from the data generating
model, we expect to see a small shortfall, with \(\lambda_2<\lambda_1\)
and \(\lambda_3<\lambda_1\). The difference, \(\lambda_2-\lambda_3\), is
statistically indistinguishable from zero in this experiment, showing
that \(J_1=5000\) particles is adequate for \(\PALV\) on a single city.

\(E_4\) and \(E_5\) demonstrate the positive bias of \(\PALL\) both at a
usual number of particles and for an intensive calculation that may not
be possible on larger problems. The best estimates of this bias are
\(\lambda_4-\lambda_3\) and \(\lambda_5-\lambda_3\), since \(\PALL\) and
\(\PALV\) target the same quantity in the limit as \(J\to\infty\).
Comparison of \(E_5\) and \(E_3\) shows that, even with a large number
of particles and a low-dimensional dynamic model, \(\PALL\) provides an
over-stated log-likelihood estimate.

Comparing \(E_4\) and \(E_5\), we see that the smaller number of
particles for \(E_4\) leads to a higher log-likelihood estimate. This
would not happen in a situation where the likelihood estimate is
unbiased with finite variance: in that situation, a higher number of
particles is expected to lead to a higher log-likelihood estimate due to
reduced negative bias resulting from Jensen's inequality.

Experiments \(E_6\) and \(E_7\) introduce the actual data, while still
restricting to a single spatial unit. We see that \(\PALL\) reports a
higher log-likelihood than BPF. At face value, this could be because the
PAL approximation is a superior model to the partially observed Markov
process model implemented by BPF. Or, it could be because BPF suffers
from heavy negative bias due to high Monte Carlo error combined with
Jensen's inequality. The latter is not the case due to BPF's empirically
low Monte Carlo error. We have just discovered in \(E_1\)--\(E_5\) that
\(\PALL\) reports an over-stated log-likelihood when the truth is known,
so the most plausible explanation of \(E_6\) and \(E_7\) is simply that
the same phenomenon occurs on the data.

Experiments \(E_8\)--\(E_{15}\) investigate a 40 unit system. For
\(E_8\), \(E_9\) and \(E_{10}\), we simulated from a model with the
coupling parameter between towns set to zero. That was done to study a
situation where a block particle filter gives a consistent and
low-variance estimate of the exact log-likelihood, calculated as
\(\lambda_8\). We see the same story as the single-unit case, where the
positive bias of \(\PALL\) is estimated by
\((\lambda_{10}-\lambda_{9})/40=\) 80.55 per unit. This bias is large
enough that we also obtain \(\lambda_{10}>\lambda_8\), with the
difference being \((\lambda_{10}-\lambda_{8})/40=\) 34.69 per unit. We
see that the \(U=40\) results scale approximately linearly compared to
\(U=1\), providing us with supporting evidence for our hypothesis H2.

Experiments \(E_{11}\)--\(E_{15}\) consider coupled models for all 40
cities in the full, real dataset. Here, \(E_{11}\) uses the same choice
of shared and unit-specific parameters as model C in Table~3 of WWR,
re-optimized using iterated BPF to account for differences between this
model and the model of WWR. Experiments \(E_{12}\)--\(E_{15}\) directly
use the parameters provided by WWR for their model A. The choice of
model A for PAL was motivated by computational convenience, since this
simpler case is sufficient to demonstrate the positive bias in the
lookahead method for the 40-unit dataset.

A comparison of \(E_{14}\) and \(E_{15}\) shows the bias of \(\PALL\)
using the code and parameter values of WWR, changing only the number of
particles. This provides the most direct evidence for H1, since the
decrease of the log-likelihood by \(\lambda_{14}-\lambda_{15}=\) -833.83
log units when moving from \(J_1=5\times 10^3\) to \(J_2=10^5\) suggests
strongly that the log-likelihood estimate with \(J_1=5\times 10^3\) is
over-optimistic.

Based on the comparison of \(E_4\) and \(E_5\), \(\PALL\) does not
report an accurate log-likelihood even on a single unit for \(J=10^5\),
and \(40\) units is a considerably harder problem. Further, we know from
\(E_8\)--\(E_{10}\) that \(\PALL\) can over-state the true
log-likelihood for the measles model with 40 units in a situation where
the truth is known. The evidence suggests that neither \(E_{14}\) nor
\(E_{15}\) is a reliable estimator of either the PAL log-likelihood or
the exact, unknown, log-likelihood that PAL approximates for the data.
The performance of \(\PALV\) on the 40-unit real dataset fails to
outperform the ARMA benchmark, further indicating that applying PAL to
high-dimensional data remains challenging.

\section{Some theoretical considerations for lookahead PAL}
\label{sec:theory}

Since vanilla particle filter algorithms are unbiased for the
likelihood, it might be reasonable to expect the PAL-SMC algorithm to be
unbiased for the PAL likelihood, but this is not true for the lookahead
PAL filter used by WWR. This is a property of the lookahead part of the
algorithm, developed by \citet{rimella23}, rather than the PAL
approximation. Therefore, for the remainder of this section, we consider
the simpler lookahead filter of \citet{rimella23}.

Briefly, the vanilla particle filter is unbiased for the likelihood
(and, therefore, negatively biased for the log-likelihood) because the
self-normalization denominator in the resampling step happens to
coincide with the conditional likelihood estimate on the numerator,
leading to a fortuitous cancellation. Self-normalization does not always
lead to unbiased likelihood estimates, as we can see from the following
example.

Let \(X\) take values \(\{0,1\}\) with equal probability, and let
\(Y=X\) with probability 1. Suppose a single data point, \(Y=1\).
Suppose also an independent sample of \(J\) particles, \(x_{1:J}\), each
with distribution matching \(X\). Now, resample these particle with
probability
\(p_j = (1-\epsilon) x_j/[\sum_j x_j] + \epsilon(1-x_j)/[\sum_j 1-x_j]\)
so that, on average, a fraction \((1-\epsilon)\) of the resampled
particles have value 1. Take \(\epsilon \ll 1/J\), so that most
resampled particle swarms contain no particles with value 0. Most
particle swarms resulting from resampling will have \(x_j = 1\) for all
\(j\), with the proper resampled weight \(w_j = 1/(J\, p_j)\) being
approximately \(1/(2J)\) for all \(j\). Rare swarms will have a particle
with massive weight, approximately \(1/(2J\epsilon)\). Under
self-normalization, particle swarms with a massive weight will estimate
the likelihood to be approximately zero, and particle swarms with
\(x_j=1\) for all \(j\) will estimate the likelihood to be 1. By setting
\(\epsilon\) arbitrarily small, we can get an estimate whose expectation
approaches 1 since with high probability we see only resampled swarms
where every particle has value 1. If we take a different limit, with
\(J \to \infty\), the bias will go away asymptotically, but here we
consider the case with fixed \(J\) and \(\epsilon \to 0\).

Importantly, the bias on the likelihood estimate in this example is
positive. As mentioned earlier, a suboptimal forecast generally gives,
on average, a negative bias on the conditional log-likelihood estimate,
since log-likelihood is a proper scoring rule. This justifies assessing
filters on their log-likelihood estimate in a similar way that one does
for parameters in likelihood-based inference. A filter with a high
log-likelihood estimate on simulated data from the target model is
validated as a good likelihood approximation. However, this does not
necessarily apply to algorithms that look at future observations. When
implementing lookahead algorithms, if you want the log-likelihood
estimate to be conservative, you have to be extra careful to consider
the bias. For unbiased likelihood estimates, the negative bias on the
log-likelihood is a direct consequence of variance, and among such
estimates it is reasonable to prefer a filter approximation with the
highest log-likelihood estimate. For positively biased estimates, that
is inappropriate.

\section{Conclusion}
\label{sec:conclusion}

The results in this article reinforce the investigation by
\citet{hao24-arxiv} and lead to the conclusion that there is not
currently a strong case for using PAL on the epidemiological models used
by WWR to motivate its development. Simpler particle filter methods
apply to arbitrary Markov process models, whereas PAL is limited to a
specific class of discrete-state Markov process models. Basic particle
filters, and their block particle filter extensions, have the
plug-and-play property \citep{breto09,he10}. Iterated block particle
filters can be effective tools for data analysis of spatiotemporal
epidemiological data \citep{wheeler24,li24}. Certain automatically
differentiable particle filter algorithms inherit this convenient
plug-and-play property while providing derivative estimates with low
bias and variance \citep{tan24}, and these may in future be extended to
block particle filters.

PAL may potentially lead to dramatic computational improvements over
particle filters some situations. However, WWR's overdispersed
generalization of PAL also involves a block particle filter component,
at which point it shares many of the properties of particle filters. The
results of WWR, together with various other authors
\citep{stocks18,he10,li24}, show that overdispersion is frequently
necessary for a dynamic model to provide an adequate statistical
description of epidemiological data.

\section*{Reproducibility}

Code and data to generate this article are available at
https://zenodo.org/records/15817139.

\section*{Acknowledgments}

Michael Whitehouse, Nick Whiteley and Lorenzo Rimella provided
constructive feedback on an earlier version of this article.

\bibliography{bib-pal-measles.bib}

\end{document}